\documentclass{article}
\usepackage{graphicx} 
\usepackage{amsmath,amssymb}
\usepackage{optidef}
\usepackage[style=authoryear, backend=biber, natbib=true,
uniquelist=false, uniquename=false, dashed=false,
sorting=nyt, maxbibnames=99, minbibnames=99]{biblatex}

\DeclareNameAlias{sortname}{family-given}
\DeclareNameAlias{default}{family-given}

\renewbibmacro*{date+extradate}{}
\renewbibmacro*{labeldate+extradate}{}

\DeclareFieldFormat[article]{title}{#1}
\DeclareFieldFormat[article]{journaltitle}{\mkbibemph{#1}}
\DeclareFieldFormat[book]{title}{\mkbibemph{#1}}
\DeclareFieldFormat[misc]{title}{#1}
\DeclareFieldFormat[article]{volume}{#1}
\DeclareFieldFormat[article]{number}{\mkbibparens{#1}}
\DeclareFieldFormat{pages}{#1}
\DeclareFieldFormat{doi}{\ifhyperref{\href{https://doi.org/#1}{\nolinkurl{https://doi.org/#1}}}{\nolinkurl{https://doi.org/#1}}}

\DeclareBibliographyDriver{article}{%
	\usebibmacro{bibindex}%
	\usebibmacro{begentry}%
	\usebibmacro{author}%
	\addperiod\space
	\mkbibparens{\printfield{year}}%
	\addperiod\space
	\printfield{title}%
	\addperiod\space
	\printfield{journaltitle}%
	\addcomma\space
	\printfield{volume}%
	\printfield{number}%
	\addcomma\space
	\printfield{pages}%
	\iffieldundef{doi}{}{\addperiod\space\printfield[doi]{doi}}%
	\addperiod
	\usebibmacro{finentry}%
}

\DeclareBibliographyDriver{book}{%
	\usebibmacro{bibindex}%
	\usebibmacro{begentry}%
	\usebibmacro{author}%
	\addperiod\space
	\mkbibparens{\printfield{year}}%
	\addperiod\space
	\printfield{title}%
	\addperiod\space
	\printlist{publisher}%
	\addperiod
	\usebibmacro{finentry}%
}

\DeclareBibliographyDriver{misc}{%
	\usebibmacro{bibindex}%
	\usebibmacro{begentry}%
	\usebibmacro{author}%
	\addperiod\space
	\mkbibparens{\printfield{year}}%
	\addperiod\space
	\printfield{title}%
	\addperiod\space
	\printfield{howpublished}%
	\iffieldundef{doi}{}{\addperiod\space\printfield[doi]{doi}}%
	\addperiod
	\usebibmacro{finentry}%
}

\DeclareBibliographyDriver{online}{%
	\usebibmacro{bibindex}%
	\usebibmacro{begentry}%
	\usebibmacro{author}%
	\addperiod\space
	\mkbibparens{\printfield{year}}%
	\addperiod\space
	\printfield{title}%
	\addperiod\space
	\iffieldundef{type}{}{\printfield{type}\iffieldundef{number}{}{\addspace No.\addspace\printfield{number}}}%
	\iffieldundef{doi}{}{\addperiod\space\printfield[doi]{doi}}%
	\addperiod
	\usebibmacro{finentry}%
}

\AtEveryBibitem{%
	\clearfield{url}%
	\clearfield{urldate}%
	\clearfield{eprint}%
	\clearfield{eprinttype}%
	\clearfield{note}%
	\clearfield{issn}%
	\clearfield{isbn}%
	\clearfield{abstract}%
	\clearfield{keywords}%
	\clearfield{month}%
	\clearfield{edition}%
	\clearlist{language}%
	\clearlist{location}%
	\clearfield{address}%
	\clearfield{series}%
	\clearfield{pagetotal}%
	\clearfield{shortjournal}%
	\clearfield{pubstate}%
}
\DeclareCiteCommand{\cite}
{}
{\bibhyperref{\printfield{year}}}
{\multicitedelim}
{}
\let\citeyear\cite
\usepackage[toc,page]{appendix}
\usepackage{mathtools}
\usepackage{pgfplotstable}
\usepackage{siunitx}
\usepackage{booktabs}
\usepackage{float}
\usepackage{bm}
\usepackage{authblk}
\usepackage{hyperref}

\newcommand\numberthis{\addtocounter{equation}{1}\tag{\theequation}}

\begin{document}
	
	\title{On the Structure of Risk Contribution: A Leave-One-Out Decomposition into Inherent and Correlation Risk}

	\author[1]{Nolan Alexander}
	\author[2]{Frank Fabozzi}
	\affil[1]{University of Virginia's School of Engineering and Applied Sciences, Charlottesville, VA. Contact: nka5we@virginia.edu}
	\affil[2]{Johns Hopkins University's Carey Business School, Baltimore, MD. Contact: fabozzi321@aol.com}

	\date{}

	\maketitle
	
	\begin{abstract}
		\noindent Risk Contribution (RC) is a widely used risk attribution metric that assigns each position a share of portfolio volatility, with contributions that sum to total portfolio volatility. This paper proposes a leave-one-out decomposition of RC into two interpretable components at the position level: an inherent term (position's own scaled variance) and a correlation term (covariance with the other holdings). Unlike the X-sigma-rho framework, which uses the position's correlation with the full portfolio, ICD uses covariance with the other holdings, yielding qualitatively different diagnostic signals when a position is large relative to the portfolio. The contribution is the interpretive framing and an exact identity linking the components to incremental volatility (iVol), which allows the discrete condition under which a position reduces portfolio volatility to be stated in closed form. That condition differs from the marginal condition and the two can disagree for over-hedged or levered positions. The decomposition also inherits RC's strict additivity. Empirical illustrations on three multi-asset portfolios show the components shift differently across regimes, distinguishing stand-alone volatility spikes from correlation breakdown, and demonstrate value for risk reporting, stress testing, and attribution.
	\end{abstract}
	

	\bigskip
	\noindent \textbf{Key messages:}
	\begin{itemize}
		\item RC separates into an inherent variance term and a correlation-with-others term.
		\item An exact identity links the two components to incremental volatility.
		\item Discrete and marginal hedging conditions can disagree for large positions.
		\item Crisis attribution shows the components shift differently across market regimes.
	\end{itemize}

	\section{Introduction}
	
	Risk attribution measures how each position contributes to total portfolio volatility. Incremental Volatility (iVol) provides an intuitive leave-one-out measurement but is non-additive: individual contributions do not sum to total portfolio volatility. Risk Contribution (RC) is additive and widely used in optimization and reporting systems, but does not reveal whether risk derives from position volatility or correlation structure.
	
	Both report a single number per position, but positions risky in isolation require different responses than positions risky due to correlation. Aggregate RC does not separate these cases.
	
	We decompose RC into an inherent term (position's own scaled variance) and a correlation term (covariance with the other holdings), called the \textit{Inherent and Correlation Decomposition} (ICD). The contribution is the specific split, its interpretive framing, and an exact identity linking the two components to iVol.
	
	The decomposition clarifies when positions function as effective hedges, how risk evolves across regimes, and supports hierarchical risk reporting. Empirical illustrations demonstrate its diagnostic value for portfolio management. Unlike many theoretical constructs, the decomposition can be implemented with existing portfolio data and covariance estimates.
	
	The risk-budgeting literature, including Litterman (\cite{littermanHotSpotsHedges1996}), Qian (\cite{qianFinancialInterpretationRisk2005d}), and Roncalli (\cite{roncalliIntroductionRiskParity2014}), works with RC as a single number per position. In that literature, RC serves primarily as an optimization input or equalization target, which are contexts where the internal structure of a single RC value is not relevant. The only prior decomposition of RC is the multiplicative X-sigma-rho framework of Menchero and Davis (\cite{mencheroRiskContributionExposure2011}), which breaks RC into exposure, volatility, and correlation factors. That decomposition answers the question of what the multiplicative drivers of RC are, rather than how much of RC derives from stand-alone variance versus covariance with the other holdings. Because a prominent multiplicative decomposition already existed, and the literature's framing was optimization rather than diagnosis, the question of an additive split into own-variance and interaction terms appears not to have been pursued. The additive inherent-correlation decomposition proposed here separates stand-alone variance from interaction variance at the position level, and links exactly to iVol as developed in Section~\ref{sec:ivol_connection}. Comparison to X-sigma-rho appears in Section~\ref{sec:diagnostic_insight}.
	
	The paper makes two principal contributions. First, it introduces the inherent-correlation decomposition of RC at the position level and derives an exact identity linking the two components to incremental volatility. That identity yields a closed-form discrete condition for when a position reduces portfolio risk, which differs from the marginal condition $\text{RC}(a) < 0$ used implicitly in existing practice. Second, the paper details the diagnostic value of the decomposition for hedging analysis, stress-period attribution, and hierarchical risk reporting, together with the granularity effects that constrain how the components should be compared across aggregation levels.

	Supporting these contributions, we provide two methods for time-varying risk analysis across historical market regimes. Appendix~\ref{app:convergence} examines convergence of the decomposition's components under a stylized simulation.
	
	Section~\ref{sec:lit_review} reviews related work, Section~\ref{sec:rc_loo} derives the leave-one-out representation, Section~\ref{sec:properties} develops its properties, Section~\ref{sec:empirical} applies it to multi-asset portfolios, and Section~\ref{sec:conclusion} concludes.
	
	\section{Review of Risk Models} \label{sec:lit_review}
	
	Markowitz (\cite{markowitzPortfolioSelection1952}) introduced Modern Portfolio Theory, measuring portfolio risk as a function of individual volatilities and covariance structure. Diversification arises from imperfect correlations, making accurate covariance estimation central to risk measurement.
	
	Risk management practice introduced measures capturing extreme losses: Value at Risk (VaR), from the RiskMetrics framework (\cite{longerstaeyRiskMetricsTechnicalDocument1996}), and Conditional Value at Risk (CVaR). These are widely used but do not identify individual asset contributions. Incremental volatility (iVol) addresses this by measuring the change in portfolio volatility when a specific asset is removed, providing a leave-one-out interpretation of risk contribution (Dowd (\cite{dowdValueRiskNew1998})).
	
	However, iVol is non-additive: individual contributions do not sum to total portfolio volatility (see Appendix~\ref{app:ivol_nonadditivity} for a proof holding for all weight vectors). Mignacca and Fusai (\cite{mignaccaIncrementalVolatilityRelated2023}) improve computational efficiency for iVol, but non-additivity is inherent to the leave-one-out construction.
	
	Risk Contribution (RC), formalized by Grinold and Kahn (\cite{grinoldActivePortfolioManagement1999}), derives each asset's contribution from its covariance with the portfolio return, which allows contributions to sum to total portfolio volatility. Tasche (\cite{tascheRiskContributionsPerformance1999}) shows that this gradient-based definition is the only one suitable for performance measurement, and Denault (\cite{denaultCoherentAllocationRisk2001}) provides an axiomatic justification via cooperative game theory. Litterman (\cite{littermanHotSpotsHedges1996}) uses the analogous marginal contributions to identify concentrations of risk and hedge candidates. Qian (\cite{qianFinancialInterpretationRisk2005d}) shows that these marginal contributions can be interpreted as financial contributions that add up to total portfolio risk, and Roncalli (\cite{roncalliIntroductionRiskParity2014}) provides a comprehensive treatment of risk budgeting built on the same identity. These works report and use RC as a single number per position. Menchero and Davis (\cite{mencheroRiskContributionExposure2011}) provide a multiplicative decomposition of RC into exposure, volatility, and correlation, known as X-sigma-rho, which decomposes each RC into three factors. Boudt, Peterson, and Croux (\cite{boudtEstimationDecompositionDownside2008}) extend component risk decomposition to non-normal return distributions. Meucci (\cite{meucciRiskContributionsGeneric2006}) derives risk contributions for generic non-linear portfolios, and Roncalli and Weisang (\cite{roncalliRiskParityPortfolios2016a}) survey the broader risk-parity landscape.

	The additive inherent-correlation decomposition of RC at the position level proposed here has a different structure. It splits each RC into two additive terms that separate stand-alone variance from interaction variance, and it links exactly to iVol via a closed-form identity used later to state the discrete condition under which a position reduces portfolio risk. The comparison to X-sigma-rho is developed in Section~\ref{sec:diagnostic_insight}.

	The specific additive split has not, to our knowledge, appeared in prior work. The algebraic step of separating the own-variance term from the interaction term in $\operatorname{cov}(r_a, r_p) = w_a\sigma_a^2 + \operatorname{cov}(r_a, r_p - w_ar_a)$ is implicit in many derivations of the covariance matrix row, but it does not appear to have been substituted back into the RC formula with the resulting terms named and given an economic interpretation. The closest technical precursor is Mignacca and Fusai (\cite{mignaccaIncrementalVolatilityRelated2023}), who work with the leave-one-out variance $\operatorname{Var}(r_p - w_ar_a)$ to derive efficient formulas for incremental volatility. Their analysis operates on the same mathematical objects and requires the same covariance expansion. Their focus is on computing iVol efficiently rather than decomposing RC into interpretable components, however, so the step of splitting $\operatorname{cov}(r_a, r_p)$ and naming the resulting additive terms was not taken. The iVol identity, the discrete hedging condition, and the granularity analysis are what distinguish the decomposition from an algebraic rearrangement of existing expressions.
	
	Both RC and iVol depend on covariance estimation quality; errors can significantly affect risk measurement and optimization (Chopra and Ziemba (\cite{chopraEffectErrorsMeans1993})). Methods include GARCH models for time-varying covariance (Engle (\cite{engleTheoreticalEmpiricalProperties2001a})) and shrinkage estimators (Ledoit and Wolf (\cite{ledoitImprovedEstimationCovariance2003})).
	
	The leave-one-out representation used in this paper builds on the risk-budgeting and covariance-estimation literature. The representation preserves the additive properties of RC and exposes a covariance term that connects RC directly to iVol, which supports the interpretive framing of the two components and the discrete hedging condition developed in later sections.
	
	\section{A Leave-One-Out Representation of RC} \label{sec:rc_loo}
	
	Additivity allows hierarchical decomposition across asset class, region, sector, or strategy. It enables a direct comparison between elements and conveys the proportion of risk any element contributes to a higher-level aggregate. Non-additive measures like iVol cannot be summed to form a decomposition of total portfolio risk.
	
	RC is strictly additive, and this section derives a leave-one-out representation of RC. Let $N$ denote the set of asset indices in the portfolio $p$, and let $a \in N$ denote an individual asset. Let $\sigma_p$ denote the volatility of the portfolio and $\sigma_{p\backslash \{a\}}$ denote the volatility of the portfolio with asset $a$ removed. The iVol of asset $a$ is then:
	
	\begin{equation} \label{eq:ivol_def}
		\text{iVol}(a) = \sigma_p - \sigma_{p\backslash \{a\}}.
	\end{equation}
	
	\noindent Let $w_a$ and $r_a$ denote the weight and return of asset $a$ respectively, and let $r_p$ denote the return of the portfolio. The risk contribution of asset $a$ is defined as its covariance with the portfolio return, weighted and normalized so that contributions sum to total portfolio volatility:
	
	\begin{equation} \label{eq:rc}
		\text{RC} (a) = \frac{ \sum_{i \in N} w_a w_i \operatorname{cov}(r_a, r_i)}{\sigma_p}.
	\end{equation}
	
	\noindent Menchero and Davis (\cite{mencheroRiskContributionExposure2011}) show that risk contribution can be reformulated as
	
	\begin{equation} \label{eq:rc_menchero}
		\text{RC} (a) = w_a \sigma_a \rho(r_a, r_p).
	\end{equation}
	
	\noindent Note that RC is defined here in units of volatility: the sum of $\text{RC}(a)$ over all assets equals the portfolio volatility $\sigma_p$. Because $w_a \sigma_a \rho(r_a, r_p) = w_a \operatorname{cov}(r_a, r_p) / \sigma_p$, the Menchero-Davis form can equivalently be written as
	
	\begin{equation} \label{eq:rc_cov}
		\text{RC}(a) = \frac{w_a \operatorname{cov}(r_a, r_p)}{\sigma_p}.
	\end{equation}
	
	Let $r_p - w_a r_a = \sum_{i \in N \backslash \{a\}} w_i r_i$ denote the return of the other holdings without renormalizing their weights. We decompose the covariance of asset $a$ with the portfolio into the sum of the covariance of that asset with all assets in the portfolio, and then rearrange terms yielding

	\begin{align*}
		\operatorname{cov}(r_a, r_p) &= \sum_{i \in N} w_i \operatorname{cov}(r_a, r_i) \\
		&= w_a \sigma_a^2 + \sum_{i \in N \backslash \{a\}} w_i \operatorname{cov}(r_a, r_i) \\
		&= w_a \sigma_a^2 + \operatorname{cov}(r_a, r_p - w_a r_a). \numberthis \label{eq:port_cov}
	\end{align*}

	\noindent Substituting eq.~\eqref{eq:port_cov} into eq.~\eqref{eq:rc_cov} gives

	\begin{equation} \label{eq:rc_loo}
		\text{RC}(a) = \frac{w_a^2 \sigma^2_a + w_a \operatorname{cov}(r_a, r_p - w_a r_a)}{ \sigma_p}.
	\end{equation}

	\noindent The first term captures the contribution from asset $a$'s own variance, and the second term captures the asset's covariance with the other holdings, taken as they are without renormalization. The risk-budgeting literature, including Litterman (\cite{littermanHotSpotsHedges1996}), Qian (\cite{qianFinancialInterpretationRisk2005d}), and Roncalli (\cite{roncalliIntroductionRiskParity2014}), works with $\text{RC}(a)$ as a single number per position and does not decompose it in this way. We work with $\operatorname{cov}(r_a, r_p - w_a r_a)$ directly to avoid the rescaling $(r_p - w_a r_a) / (1 - w_a)$ often used in leave-one-out constructions. That rescaling is undefined when $w_a = 1$ (e.g., a dollar-neutral overlay), reverses the sign of the other holdings' returns when $w_a > 1$, and implicitly requires the weights to sum to a specific total. Eq.~\eqref{eq:rc_loo} is well defined for arbitrary weights, including short and levered positions. It creates the exact connection to iVol used in Section~\ref{sec:ivol_connection} and supports the decomposition described in Section~\ref{sec:risk_decomp}.
	
	\section{Properties of the Leave-One-Out Representation} \label{sec:properties}
	
	This section develops four properties of the leave-one-out representation: strict additivity, an exact link to iVol, the decomposition into inherent and correlation components with its associated hedging condition and granularity caveat, and diagnostic insight relative to aggregate RC and to X-sigma-rho. Statistical behavior and time-varying analysis follow.
	
	\subsection{Strict Additivity}
	
	Unlike RC, iVol is not strictly additive: the sum of individual iVols does not equal portfolio volatility. From eq.~\eqref{eq:ivol_def},
	\begin{equation} \label{eq:ivol_nonadditivity}
		\sum_{a \in N} \text{iVol}(a) = |N|\,\sigma_p - \sum_{a \in N} \sigma_{p\backslash \{a\}}.
	\end{equation}
	Appendix~\ref{app:ivol_nonadditivity} shows that $\sum_{a \in N} \text{iVol}(a) \leq \sigma_p$ holds for all weight vectors, with strict inequality for any non-degenerate portfolio. The sum of iVols therefore systematically under-attributes total portfolio volatility. Non-additivity of iVol itself is well known in the risk-budgeting literature.
	
	Unlike iVol, the leave-one-out ICD representation of RC is strictly additive, inheriting this property from RC itself. Applying eq.~\eqref{eq:port_cov} to each term in the sum gives
	
	\begin{align*}
		\sum_{i\in N} \text{RC}(i)
		&= \frac{\displaystyle\sum_{i\in N} \left(w_i^2\sigma^2_i + w_i\operatorname{cov}(r_i, r_p - w_i r_i)\right)}{\sigma_p} \\
		&= \frac{\displaystyle\sum_{i\in N} w_i \operatorname{cov}(r_i, r_p)}{\sigma_p} \\
		&= \frac{\displaystyle\sum_{i\in N}\sum_{j\in N} w_i w_j \operatorname{cov}(r_i, r_j)}{\sigma_p} \\
		&= \sigma_p.
	\end{align*}
	
	\noindent The same argument applies to any subset $N_{\text{sub}} \subseteq N$, so RC sums consistently at any aggregation level, enabling attribution across asset class, region, sector, and position.
	
	\subsection{Connection to iVol} \label{sec:ivol_connection}
	The inherent and correlation components determine iVol exactly, given portfolio volatility and the position's weight. Expanding the variance of $r_p - w_a r_a$ gives

	\begin{equation} \label{eq:var_expand}
		\operatorname{Var}(r_p - w_a r_a) = \sigma_p^2 + w_a^2 \sigma_a^2 - 2 w_a \operatorname{cov}(r_a, r_p),
	\end{equation}

	\noindent and substituting eq.~\eqref{eq:port_cov} for $\operatorname{cov}(r_a, r_p)$ yields

	\begin{equation} \label{eq:ivol_exact}
		\operatorname{Var}(r_p - w_a r_a) = \sigma_p \bigl( \sigma_p - \text{RC}_{\text{inh}}(a) - 2\, \text{RC}_{\text{corr}}(a) \bigr).
	\end{equation}

	\noindent The components enter eq.~\eqref{eq:ivol_exact} with different weights: inherent once, correlation twice. This asymmetry underlies the discrete hedging condition (Section~\ref{sec:risk_decomp}) and distinguishes it from the marginal condition $\text{RC}(a) < 0$. Unlike iVol, the components retain strict additivity.
	
	\subsection{Decomposition of Risk} \label{sec:risk_decomp}

	RC decomposes into inherent and correlation components:

	\begin{equation} \label{eq:rc_decomp}
		\begin{aligned}
			\text{RC}(a) &= \text{RC}_{\text{inh}} (a) + \text{RC}_{\text{corr}} (a), \\
			\text{RC}_{\text{inh}} (a) &= \frac{ w_a^2 \sigma^2_a}{\sigma_p} \ge 0, \\
			\text{RC}_{\text{corr}} (a) &= \frac{w_a \operatorname{cov}(r_a, r_p - w_a r_a)}{\sigma_p}.
		\end{aligned}
	\end{equation}

	\noindent The inherent component is quadratic in $w_a$, always non-negative, and depends on portfolio volatility through the denominator. The correlation component can be negative for short positions even when the underlying covariance is positive; Staum (\cite{staumNonnegativeRiskComponents2015}) studies approaches for attributing risk when components can be negative. Aggregate RC does not distinguish these effects.

	\paragraph{Discrete versus marginal hedging condition.}
	A position may reduce total portfolio risk in two distinct senses. In the marginal sense, increasing the exposure to $a$ lowers $\sigma_p$. The corresponding condition is $\text{RC}(a) < 0$, or equivalently $-\text{RC}_{\text{corr}}(a) > \text{RC}_{\text{inh}}(a)$. In the discrete sense, holding the position lowers $\sigma_p$ relative to not holding it (with other holdings unchanged). Using eq.~\eqref{eq:ivol_exact}, the position reduces risk in the discrete sense if and only if $\operatorname{Var}(r_p - w_a r_a) > \sigma_p^2$, which simplifies to

	\begin{equation} \label{eq:discrete_hedge}
		\text{RC}_{\text{corr}}(a) < -\tfrac{1}{2}\,\text{RC}_{\text{inh}}(a).
	\end{equation}

	\noindent The discrete threshold is weaker than the marginal threshold, because the correlation component enters eq.~\eqref{eq:ivol_exact} with weight two while the inherent component enters with weight one. An over-hedged position illustrates the gap. Consider an asset $X$ with unit volatility hedged by $h$ units of $H$ where $\sigma_H = 1$ and $\rho_{XH} = -1$. The minimum-variance hedge is $h^\star = 1$. For $h = 1.5$, $\sigma_p = 0.5$ compared with $1$ without the hedge, so the position clearly reduces risk. Yet $\text{RC}(H) = 1.5 > 0$ (with $\text{RC}_{\text{inh}} = 4.5$ and $\text{RC}_{\text{corr}} = -3$), so the marginal condition is not satisfied. The discrete condition, $-3 < -\tfrac{1}{2}(4.5) = -2.25$, is. The X-sigma-rho decomposition of the same example gives $\operatorname{cov}(r_H, r_p) = w_X\operatorname{cov}(r_H, r_X) + w_H\sigma_H^2 = (1)(-1) + (1.5)(1) = 0.5$, so $\rho(r_H, r_p) = 0.5/(\sigma_H\,\sigma_p) = 1.0$. The correlation factor equals $+1$ because $\rho(r_H, r_p)$ measures the correlation of $H$ with the full portfolio including $H$ itself; when $H$ is large, its own variance dominates and forces this factor toward $+1$ regardless of the hedging role. The ICD correlation component $\text{RC}_{\text{corr}}(H) = -3$ gives the opposite diagnostic signal because it measures $H$'s covariance with the other holdings $r_p - w_H r_H$, which is the quantity that determines whether removing $H$ raises or lowers $\sigma_p$. In interpreting the tornado chart in Section~\ref{sec:empirical} and elsewhere, this distinction is important. A position may have positive RC and still reduce portfolio volatility relative to not holding it.

	\paragraph{Granularity.}
	The decomposition depends on how positions are booked. If a single position with weight $w_a$ is split into $n$ identical tranches of weight $w_a / n$, total RC is unchanged but $\text{RC}_{\text{inh}}$ falls by a factor of $n$ (because it is quadratic in weight), while $\text{RC}_{\text{corr}}$ rises by the same amount so that the sum is preserved. The same effect appears in aggregation: the inherent component computed at the asset-class level differs from the sum of constituent inherent components. For risk reporting systems, the practical response is to fix the granularity at the level at which trades are executed, typically the individual position level, and apply the decomposition consistently at that level. At a fixed level, the ratio $\text{RC}_{\text{inh}}(a) / \text{RC}(a)$ provides a stable measure of how much of a position's risk contribution is attributable to its own volatility versus its interaction with other holdings, and comparisons of this ratio across positions are consistent. Cross-level comparisons, such as comparing inherent fractions at the position level with the same fractions at the asset-class level, should not be made without accounting for the aggregation effect described above.

	\subsection{Diagnostic Insight of the Decomposition} \label{sec:diagnostic_insight}
	
	First, the decomposition's value is diagnostic: two positions with similar RC may have very different structures. One may be intrinsically volatile (calling for exposure reduction); another may be risky because of its correlation with other holdings (calling for diversification). Aggregate RC does not make this distinction.
	
	Second, the discrete condition eq.~\eqref{eq:discrete_hedge} clarifies hedging: positions with positive RC can still reduce portfolio volatility. Distinguishing discrete from marginal hedging separates apparent from effective hedges.
	
	Third, tracking components separately across time distinguishes whether a crisis risk rise is driven by increased stand-alone volatility or correlation breakdown, with different implications for stress testing and post-mortem analysis.
	
	Fourth, the components aggregate consistently across assets, sectors, and asset classes, allowing hierarchical reporting to specify whether group-level risk is primarily inherent or correlation-driven. Granularity must be fixed (Section~\ref{sec:risk_decomp}); single-level comparisons are consistent, cross-level comparisons are not.
	
	Menchero and Davis's X-sigma-rho framework provides a related but different lens on RC by decomposing it multiplicatively into exposure, volatility, and correlation. That decomposition answers what a position's RC is composed of in terms of its weight, its volatility, and its correlation with the portfolio. The additive decomposition here answers a different question: how much of a position's RC is attributable to its stand-alone variance and how much to its interaction with the other holdings. The key structural difference is the correlation object: X-sigma-rho uses $\rho(r_a, r_p)$, the correlation of the position with the full portfolio, while ICD uses $\operatorname{cov}(r_a, r_p - w_a r_a)$, the covariance with the other holdings. These can give qualitatively different signals when a position is large relative to the portfolio, as the over-hedged example in Section~\ref{sec:risk_decomp} illustrates: $\rho(r_H, r_p) = 1.0$ while $\text{RC}_{\text{corr}}(H) = -3$, reaching opposite diagnostic conclusions. The same contrast appears empirically in Section~\ref{sec:empirical}.

	\subsection{Statistical Behavior}

	The statistical behavior of RC and its components is inherited from the covariance estimator. Options include shrinkage (Ledoit and Wolf (\cite{ledoitImprovedEstimationCovariance2003})) and robust estimators; the sample covariance is generally least stable. McNeil, Frey, and Embrechts (\cite{mcneilQuantitativeRiskManagement2015}) provide a comprehensive treatment of dependence estimation, and Olschewsky, Frahm, and Fr\"{o}hlich (\cite{olschewskyFiniteDifferenceMethods2016}) derive finite-difference standard errors for covariance-based risk measures. A systematic comparison is beyond this paper's scope; the empirical section uses exponentially weighted covariance, and we recommend evaluating alternatives in practice. Appendix~\ref{app:convergence} shows via a stylized simulation that the decomposition does not introduce additional estimation instability relative to aggregate RC; confidence bands are recommended for portfolios with limited history.
	
	\subsection{Time-varying Risk Analysis}
	
	The leave-one-out representation supports time-varying analysis. Let $\text{RC}_{s,t}(a)$ denote RC with position weights at date $s$ and covariance through date $t$, and $L$ the lookback window. The two approaches are

	\begin{equation} \label{eq:time_varying_methods}
		\begin{aligned}
			\text{Method 1:} \quad & \text{RC}_{T,t}(a), & t \in [L, T], \; a \in N, \\
			\text{Method 2:} \quad & \text{RC}_{t,t}(a), & t \in [L, T], \; a \in N.
		\end{aligned}
	\end{equation}

	\noindent In Method 1, the position date is fixed at the current date $T$ and the covariance date $t$ varies. In Method 2, both dates advance together.

	Method 1 shows what the current portfolio's risk would have been historically, useful for comparing current levels to stressed periods. Method 2 tracks how risk evolved as positions changed, useful for post-mortem analysis.
	
	\section{An Application to Portfolios} \label{sec:empirical}
	
	We apply the decomposition to three portfolios of futures and FX instruments (Bloomberg, 1990--Aug 2025; each series enters when available): (i) a long-short (LS) portfolio representative of hedge fund strategies, (ii) an equal-weight (Eq) portfolio, and (iii) an inverse-volatility (IV) portfolio, which weights positions by the reciprocal of trailing six-month volatility (naive risk parity; for equal-risk-contribution portfolios, see Maillard, Roncalli, and Teiletche (\cite{maillardPropertiesEquallyWeighted2010})). Table~\ref{tab:weight_by_asset} shows the position weights. RC values are in daily volatility percentages. Covariances use a 126-business-day rolling window with exponential weights ($\alpha = 0.99$).
	
	\begin{table}[H]
		\centering
		\caption{Portfolio Positions}
		\label{tab:weight_by_asset}
		\pgfplotstableset{
			col sep=comma,
			string type,
			columns/Asset Class/.style={
				string type,
				column type={|l|},
				column name={Asset Class}
			},
			columns/Ticker/.style={
				string type,
				column type={l|},
				column name={Ticker}
			},
			columns/LS Weight/.style={
				dec sep align={r|},
				column type/.add={}{|},
				fixed,
				precision=2,
				column name={LS Weight}
			},
			columns/Eq Weight/.style={
				dec sep align={r|},
				column type/.add={}{|},
				fixed,
				precision=2,
				column name={Eq Weight}
			},
			columns/RP Weight/.style={
				dec sep align={r|},
				column type/.add={}{|},
				fixed,
				precision=2,
				column name={IV Weight}
			},
			every head row/.style={
				before row=\hline, after row=\hline
			},
			every last row/.style={after row=\hline}
		}
		\begin{filecontents*}{tables/portfolio_weights.csv}
Asset Class,Ticker,LS Weight,Eq Weight,RP Weight
Commodity,CL1 Comdty,-0.05,0.08333333333333333,0.024433927890124458
Commodity,NG1 Comdty,0.05,0.08333333333333333,0.014660999450484347
Commodity,C 1 Comdty,-0.1,0.08333333333333333,0.04571879321654759
Equity,ES1 Index,0.05,0.08333333333333333,0.03663645553340961
Equity,CF1 Index,-0.2,0.08333333333333333,0.04603836768052227
Equity,XU1 Index,0.1,0.08333333333333333,0.045117916847760654
Fixed Income,TY1 Comdty,0.9,0.08333333333333333,0.14352452612045236
Fixed Income,RX1 Comdty,-0.9,0.08333333333333333,0.14147292697450456
Fixed Income,JB1 Comdty,0.8,0.08333333333333333,0.19525553011772506
FX,CADUSD Curncy,0.1,0.08333333333333333,0.14152144394298835
FX,EURUSD Curncy,0.15,0.08333333333333333,0.09040996396487805
FX,JPYUSD Curncy,0.1,0.08333333333333333,0.07520914826060272
\end{filecontents*}
\pgfplotstabletypeset{tables/portfolio_weights.csv}
	\end{table}
	
	\noindent Positions are organized hierarchically by asset group. Table~\ref{tab:weight_by_asset} shows pre-leverage weights, which sum to approximately $1$ signed ($3.5$ absolute). All LS positions are $10\times$ leveraged, so effective weights sum to approximately $10$ signed ($35$ absolute). Eq.~\eqref{eq:rc_loo} is well defined at any weight budget.

	All futures are rolled using the Bloomberg code N:05\_0\_R, which rolls 5 days before first notice with the ratio adjustment. The three bond futures used are TY1 (US 10-year Treasury), RX1 (German 10-year Bund), and JB1 (Japanese 10-year JGB), all quoted in price terms, so percentage returns on the rolled series are the correct economic returns without additional adjustment.
	
	\subsection{Relative Portfolio Risk on a Single Day}
	
	Table~\ref{tab:rc_by_asset} provides RC and its decomposition for each position across portfolios on August 1st, 2025.
	
	\begin{table}[H]
		\centering
		\caption{RC Decomposition by Position in daily volatility percentages, where Inh = Inherent RC and Corr = Correlation RC.}
		\label{tab:rc_by_asset}
		\footnotesize
		\setlength{\tabcolsep}{3pt}
		\pgfplotstableset{
			col sep=comma,
			string type,
			columns/asset_class/.style={
				string type,
				column type={|l},
				column name={Asset Class}
			},
			columns/ticker/.style={
				string type,
				column type={l|},
				column name={Ticker}
			},
			columns/LS RC/.style={
				column type={S[table-format=-1.2, round-mode=places, round-precision=2]},
				column name={\multicolumn{1}{c}{RC}}
			},
			columns/LS Inh/.style={
				column type={S[table-format=-1.2, round-mode=places, round-precision=2]},
				column name={\multicolumn{1}{c}{Inh}}
			},
			columns/LS Corr/.style={
				column type={S[table-format=-1.2, round-mode=places, round-precision=2]|},
				column name={\multicolumn{1}{c|}{Corr}}
			},
			columns/Eq RC/.style={
				column type={S[table-format=-1.2, round-mode=places, round-precision=2]},
				column name={\multicolumn{1}{c}{RC}}
			},
			columns/Eq Inh/.style={
				column type={S[table-format=-1.2, round-mode=places, round-precision=2]},
				column name={\multicolumn{1}{c}{Inh}}
			},
			columns/Eq Corr/.style={
				column type={S[table-format=-1.2, round-mode=places, round-precision=2]|},
				column name={\multicolumn{1}{c|}{Corr}}
			},
			columns/RP RC/.style={
				column type={S[table-format=-1.2, round-mode=places, round-precision=2]},
				column name={\multicolumn{1}{c}{RC}}
			},
			columns/RP Inh/.style={
				column type={S[table-format=-1.2, round-mode=places, round-precision=2]},
				column name={\multicolumn{1}{c}{Inh}}
			},
			columns/RP Corr/.style={
				column type={S[table-format=-1.2, round-mode=places, round-precision=2]|},
				column name={\multicolumn{1}{c|}{Corr}}
			},
			every head row/.style={
				before row={\hline & & \multicolumn{3}{c|}{Long-Short} & \multicolumn{3}{c|}{Equal Weight} & \multicolumn{3}{c|}{Inverse Volatility} \\ \hline},
				after row=\hline
			},
			every last row/.style={after row=\hline}
		}
		\begin{filecontents*}{tables/last_rc_lag_adj_data.csv}
asset_class,ticker,LS RC,LS Inh,LS Corr,Eq RC,Eq Inh,Eq Corr,RP RC,RP Inh,RP Corr
Commodity,C 1 Comdty,0.30041510237281066,0.2344175502161671,0.06599755215664357,0.3688997820427961,0.21518550783277665,0.1537142742100195,0.20087332630244226,0.14167870624682843,0.059194620055613774
Commodity,CL1 Comdty,0.596061648895019,0.21068122026962366,0.38538042862539523,1.046433557677593,0.7735861983497752,0.27284735932781784,0.0857853033342844,0.14547800436095737,-0.05969270102667296
Commodity,NG1 Comdty,0.13546944186977986,0.5320063501947081,-0.3965369083249283,2.3829348991974166,1.9534383245852185,0.42949657461219787,0.10746369910652642,0.13226011779758107,-0.024796418691054655
Equity,CF1 Index,1.033376729827005,0.8324865667628232,0.20089016306418161,0.463611752104818,0.19104696349274042,0.2725647886120775,0.13386592120015695,0.1275504572415456,0.0063154639586113235
Equity,ES1 Index,-0.31011163140477993,0.0731392695425267,-0.3832509009473066,0.4844280496396117,0.2685551631183542,0.21587288652125744,0.07663656936862173,0.11354354602761765,-0.03690697665899595
Equity,XU1 Index,-0.16459487802395642,0.1981076897445192,-0.36270256776847565,0.4294611753068386,0.18185457438635266,0.2476066009204858,0.19550686125273795,0.1166069282938864,0.07889993295885152
FX,CADUSD Curncy,0.06868585092068656,0.023069366434607944,0.045616484486078614,0.09757368462387815,0.02117671363357313,0.07639697099030504,0.37943331598009544,0.13359958595708707,0.24583373002300826
FX,EURUSD Curncy,0.3684792812362167,0.13027110199709707,0.23820817923911963,0.055588170117515975,0.053148188328108,0.002439981789408002,0.2939986442853589,0.1368431629741801,0.15715548131117874
FX,JPYUSD Curncy,0.33353865204791605,0.08713019694845722,0.24640845509945886,0.008487730754419907,0.0799818770422017,-0.07149414628778174,0.31432514762979075,0.14250660017548075,0.17181854745430994
Fixed Income,JB1 Comdty,1.0834961357534427,0.7298663726204467,0.3536297631329958,-0.033978835281893716,0.01046854386392788,-0.0444473791458216,0.1428924906896929,0.1257172896523892,0.017175201037303585
Fixed Income,RX1 Comdty,0.49796675112935684,1.6298526116036902,-1.1318858604743334,-0.04818650047601703,0.018470821451189678,-0.06665732192720669,0.1583599310209584,0.11644877768125196,0.04191115333970642
Fixed Income,TY1 Comdty,1.868004323016716,1.8468359064043927,0.021168416612323368,-0.0319860734799143,0.020929853432131242,-0.05291592691204552,0.20112187869978293,0.13580647494899148,0.06531540375079146
\end{filecontents*}
\pgfplotstabletypeset{tables/last_rc_lag_adj_data.csv}
	\end{table}
	
	\noindent Table~\ref{tab:rc_by_asset_class} provides the RC and its decomposition rolled up for each asset group. The decomposition reveals distinct risk structures across portfolios. For the long-short portfolio, fixed income is the largest RC contributor, with a large positive inherent component partially offset by a negative correlation component. Commodities and FX also contribute meaningfully, while equities contribute the least in RC terms. For the equal-weighted portfolio, commodities dominate, driven by high inherent volatility, with fixed income showing a small negative RC. For the inverse-volatility portfolio, FX carries the largest RC, followed by fixed income, equities, and commodities. The IV portfolio is inverse-volatility weighted rather than an equal-risk-contribution portfolio, so the RCs are not equal by construction. The RC values here range from roughly $0.08$ to $0.38$ across positions. Comparisons of inherent components across the three portfolios should keep in mind the granularity effect discussed in Section~\ref{sec:risk_decomp}.
	
	\begin{table}[H]
		\centering
		\caption{RC Decomposition by Asset Class in daily volatility percentages, where Inh = Inherent RC and Corr = Correlation RC.}
		\label{tab:rc_by_asset_class}
		\footnotesize
		\pgfplotstableset{
			col sep=comma,
			string type,
			columns/asset_class/.style={
				string type,
				column type={|l|},
				column name={Asset Class}
			},
			columns/LS RC/.style={
				column type={S[table-format=-1.2, round-mode=places, round-precision=2]},
				column name={\multicolumn{1}{c}{RC}}
			},
			columns/LS Inh/.style={
				column type={S[table-format=-1.2, round-mode=places, round-precision=2]},
				column name={\multicolumn{1}{c}{Inh}}
			},
			columns/LS Corr/.style={
				column type={S[table-format=-1.2, round-mode=places, round-precision=2]|},
				column name={\multicolumn{1}{c|}{Corr}}
			},
			columns/Eq RC/.style={
				column type={S[table-format=-1.2, round-mode=places, round-precision=2]},
				column name={\multicolumn{1}{c}{RC}}
			},
			columns/Eq Inh/.style={
				column type={S[table-format=-1.2, round-mode=places, round-precision=2]},
				column name={\multicolumn{1}{c}{Inh}}
			},
			columns/Eq Corr/.style={
				column type={S[table-format=-1.2, round-mode=places, round-precision=2]|},
				column name={\multicolumn{1}{c|}{Corr}}
			},
			columns/RP RC/.style={
				column type={S[table-format=-1.2, round-mode=places, round-precision=2]},
				column name={\multicolumn{1}{c}{RC}}
			},
			columns/RP Inh/.style={
				column type={S[table-format=-1.2, round-mode=places, round-precision=2]},
				column name={\multicolumn{1}{c}{Inh}}
			},
			columns/RP Corr/.style={
				column type={S[table-format=-1.2, round-mode=places, round-precision=2]|},
				column name={\multicolumn{1}{c|}{Corr}}
			},
			every head row/.style={
				before row={\hline & \multicolumn{3}{c|}{Long-Short} & \multicolumn{3}{c|}{Equal Weight} & \multicolumn{3}{c|}{Inverse Volatility} \\ \hline},
				after row=\hline
			},
			every last row/.style={after row=\hline},
			every last row/.append style={
				before row=\hline
			}
		}
		\begin{filecontents*}{tables/last_asset_class_rc_lag_adj_data.csv}
asset_class,LS RC,LS Inh,LS Corr,Eq RC,Eq Inh,Eq Corr,RP RC,RP Inh,RP Corr
Commodity,1.0319461931376095,0.9771051206804988,0.05484107245711051,3.7982682389178057,2.9422100307677703,0.8560582081500352,0.3941223287432531,0.41941682840536687,-0.025294499662113844
Equity,0.5586702203982685,1.103733526049869,-0.5450633056516007,1.3775009770512683,0.6414567009974472,0.7360442760538207,0.40600935182151665,0.35770093156304966,0.04830842025846689
FX,0.7707037842048193,0.24047066538016224,0.5302331188246571,0.16164958549581404,0.15430677900388282,0.0073428064919313085,0.9877571078952451,0.4129493491067479,0.574807758788497
Fixed Income,3.449467209899516,4.20655489062853,-0.7570876807290141,-0.11415140923782505,0.0498692187472488,-0.1640206279850738,0.5023743004104342,0.37797254228263266,0.12440175812780147
Total,5.810787407640213,6.52786420273906,-0.7170767950988473,5.223267392227064,3.787842729516349,1.4354246627107135,2.290263088870449,1.5680396513577972,0.7222234375126515
\end{filecontents*}
\pgfplotstabletypeset{tables/last_asset_class_rc_lag_adj_data.csv}
	\end{table}
	
	Table~\ref{tab:rc_comparison} compares the RC decomposition to the X-sigma-rho correlation factor $\rho(r_a, r_p)$, iVol, and maximum drawdown (MDD) for the LS portfolio. Most positions show sign agreement between XSR Corr and Corr RC. Two exceptions are RX1 and NG1, where large inherent variance pulls $\rho(r_a, r_p)$ positive despite negative covariance with the other holdings. This occurs because XSR Corr is measured against the full portfolio, while ICD uses $\operatorname{cov}(r_a, r_p - w_a r_a)$. RX1 shows RC $> 0$ but iVol $< 0$, a concrete instance of the discrete-vs-marginal distinction. MDD is a property of the back-adjusted series, included for reference only.
	
	\begin{table}[H]
		\centering
		\caption{LS Portfolio RC by Position with XSR Corr, iVol, and MDD in daily volatility percentages, where Inh.\ RC = Inherent RC, Corr.\ RC = Correlation RC, XSR Corr = X-sigma-rho correlation factor $\rho(r_a, r_p)$ (a correlation coefficient), and MDD = cumulative drawdown percentage.}
		\label{tab:rc_comparison}
		\scriptsize
		\setlength{\tabcolsep}{2pt}
		\pgfplotstableset{
			col sep=comma,
			string type,
			columns={ticker, RC, Inh. RC, Corr. RC, XSR Corr, iVol, MDD},
			columns/ticker/.style={
				string type,
				column type={|l|},
				column name={Ticker}
			},
			columns/RC/.style={
				dec sep align={r|},
				column type/.add={}{|},
				fixed,
				fixed zerofill,
				precision=2,
				column name={RC}
			},
			columns/Inh. RC/.style={
				dec sep align={r|},
				column type/.add={}{|},
				fixed,
				fixed zerofill,
				precision=2,
				column name={Inh.\ RC}
			},
			columns/Corr. RC/.style={
				dec sep align={r|},
				column type/.add={}{|},
				fixed,
				fixed zerofill,
				precision=2,
				column name={Corr.\ RC}
			},
			columns/XSR Corr/.style={
				dec sep align={r|},
				column type/.add={}{|},
				fixed,
				fixed zerofill,
				precision=2,
				column name={XSR Corr}
			},
			columns/iVol/.style={
				dec sep align={r|},
				column type/.add={}{|},
				fixed,
				fixed zerofill,
				precision=2,
				column name={iVol}
			},
			columns/MDD/.style={
				dec sep align={r|},
				column type/.add={}{|},
				fixed,
				fixed zerofill,
				precision=2,
				column name={MDD}
			},
			every head row/.style={
				before row=\hline, after row=\hline
			},
			every last row/.style={after row=\hline}
		}
		\begin{filecontents*}{tables/last_rc_comparison.csv}
ticker,RC,Inh. RC,Corr. RC,XSR Corr,iVol,MDD
CL1 Comdty,0.5960616488950189,0.21068122026962366,0.38538042862539523,0.3478845145932375,0.29433062821411227,-98.34915881958946
NG1 Comdty,0.13546944186977977,0.5320063501947081,-0.3965369083249283,0.2981828073826727,0.27485619135354455,-99.9739343298499
C 1 Comdty,0.30041510237281066,0.2344175502161671,0.06599755215664357,0.1886950009941637,0.10833132605186369,-94.99104484693208
ES1 Index,-0.31011163140477993,0.0731392695425267,-0.3832509009473066,-0.22585866265165877,-0.18391400097752392,-63.65495383767414
CF1 Index,1.0333767298270047,0.8324865667628232,0.20089016306418161,0.3829664281965139,0.46018299219005365,-67.19151530990032
XU1 Index,-0.16459487802395645,0.1981076897445192,-0.36270256776847565,-0.014970969458059376,-0.11435103981816686,-77.1389001422636
TY1 Comdty,1.868004323016716,1.8468359064043927,0.021168416612323368,0.5840963778337905,1.1341949100206257,-23.939293231416876
RX1 Comdty,0.49796675112935684,1.6298526116036902,-1.1318858604743334,0.1694375674494352,-0.27686877501623686,-25.26073890754819
JB1 Comdty,1.0834961357534425,0.7298663726204467,0.3536297631329958,0.4115602535709387,0.5218745403777716,-15.439233817701398
CADUSD Curncy,0.06868585092068656,0.023069366434607944,0.045616484486078614,0.2450987452983043,0.08050844313456998,-36.87068568798927
EURUSD Curncy,0.3684792812362167,0.13027110199709707,0.23820817923911963,0.48015586509110364,0.3723273399997611,-44.479495268139075
JPYUSD Curncy,0.3335386520479161,0.08713019694845722,0.24640845509945886,0.45951250370571695,0.2971497114334018,-53.10130421595377
\end{filecontents*}
\pgfplotstabletypeset{tables/last_rc_comparison.csv}
	\end{table}
	
\begin{figure}[H]
		\centering
		\includegraphics[width=1\textwidth]{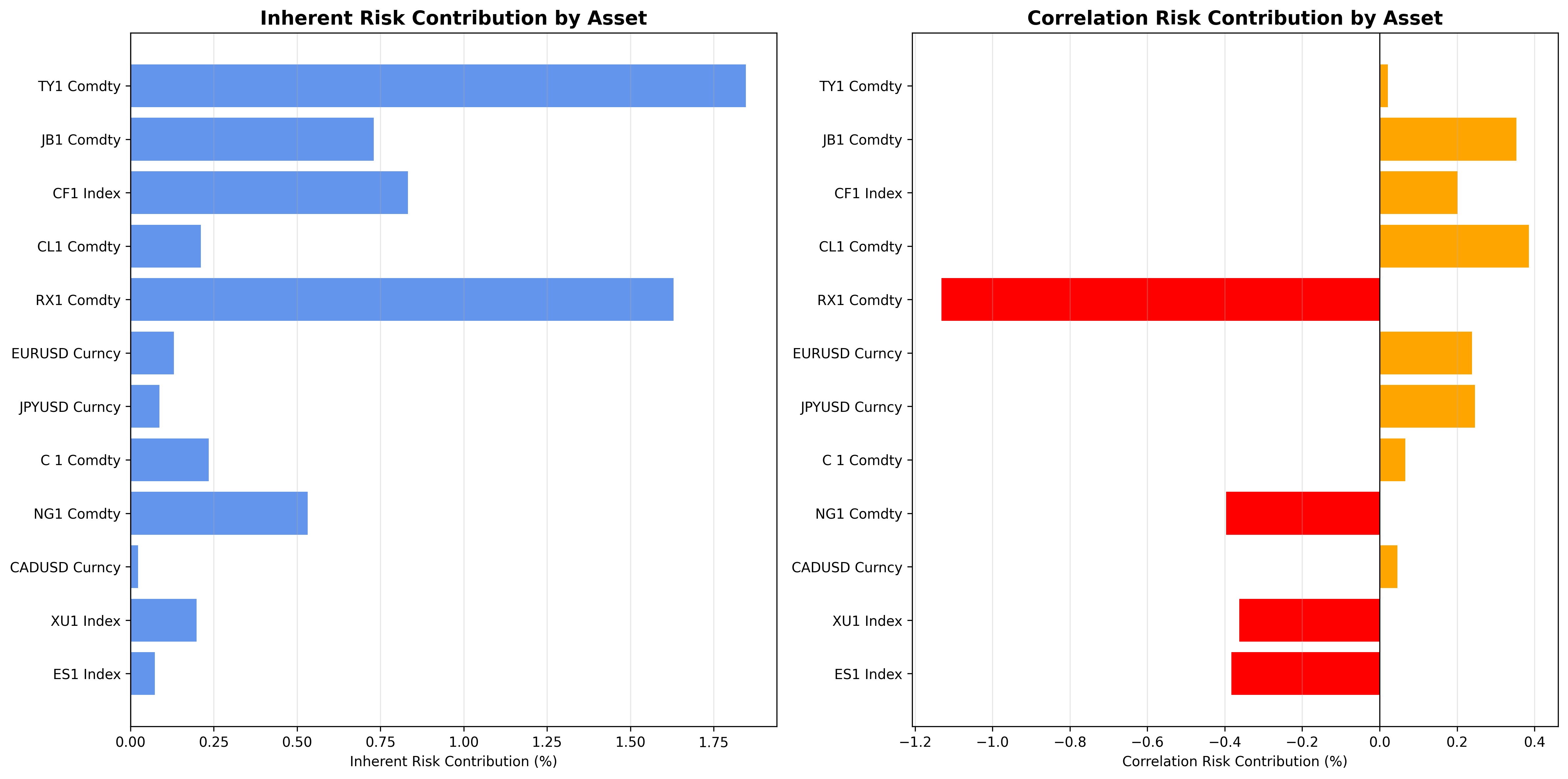}
		\caption{RC of the long-short portfolio decomposed into inherent risk and correlation risk (left and right panels, respectively), ordered by total RC, with inherent bars non-negative by construction and correlation bars potentially positive or negative.}
		\label{fig:rc_tornado}
	\end{figure}

	Figure~\ref{fig:rc_tornado} visualizes the RC decomposition for the long-short portfolio as two side-by-side bar charts, one for each component, sorted so that the two panels share the same ordering.

	Reading Table~\ref{tab:rc_comparison} in conjunction with the figure, several patterns stand out. TY1 has the largest RC, driven almost entirely by its inherent component, with a correlation component close to zero. JB1 is the second-largest RC contributor, with positive inherent and positive correlation components. CF1 also has a large RC with a large inherent component and a positive correlation component. RX1 has a large inherent component substantially offset by a large negative correlation component, with the correlation component offsetting roughly $70\%$ of the inherent component. RX1 satisfies the discrete condition in eq.~\eqref{eq:discrete_hedge} comfortably, and its iVol is clearly negative, so removing RX1 would meaningfully increase portfolio volatility despite $\text{RC}(\text{RX1}) > 0$. This is a concrete empirical instance of the over-hedged case described in Section~\ref{sec:risk_decomp}. The XSR Corr column in Table~\ref{tab:rc_comparison} allows a systematic comparison across all positions. For the majority of positions, XSR Corr and Corr.\ RC agree in sign: positions with positive Corr.\ RC also show positive XSR Corr, and vice versa. Two positions show sign disagreements: RX1 and NG1. For RX1, XSR Corr is weakly positive while Corr.\ RC is strongly negative ($-1.13$). For NG1, XSR Corr is moderately positive ($+0.30$) while Corr.\ RC is negative ($-0.40$). In both cases, the position's inherent variance is large relative to its covariance with the other holdings, so $\rho(r_a, r_p)$ is pulled toward a positive value by the position's own variance despite the hedging role the position plays. These are empirical instances of the diagnostic difference described in Section~\ref{sec:risk_decomp} and illustrated analytically with the over-hedged example. The ICD components $\text{RC}_{\text{inh}}(\text{RX1}) = 1.63$ and $\text{RC}_{\text{corr}}(\text{RX1}) = -1.13$ enable a direct check of eq.~\eqref{eq:discrete_hedge}: $-1.13 < -\tfrac{1}{2}(1.63) = -0.815$, confirming the discrete hedge, consistent with $\text{iVol}(\text{RX1}) = -0.28$. The XSR Corr value alone does not support this check without additional computation. Among the bond futures, TY1, JB1, and RX1 all carry positive RC. None satisfies the marginal hedging condition $\text{RC}(a) < 0$. ES1 and XU1 are the two positions with negative RC, and both have $\text{iVol}(a) < 0$ as well, so both act as discrete hedges in the leave-one-out sense. The FX positions all contribute positively to RC through both components. Comparing across asset groups, most of the commodity RC is inherent. Fixed-income positions carry large inherent components, with the offsetting correlation components varying widely by position, from strongly negative for RX1 to nearly zero for TY1 and positive for JB1.
	
	\subsection{Time-Varying Relative Portfolio Risk}
	This subsection applies the two time-varying methods from Section~\ref{sec:properties} to all three portfolios and identifies how the inherent and correlation components evolve across market regimes.

\begin{figure}[H]
		\centering
		\includegraphics[width=1\textwidth]{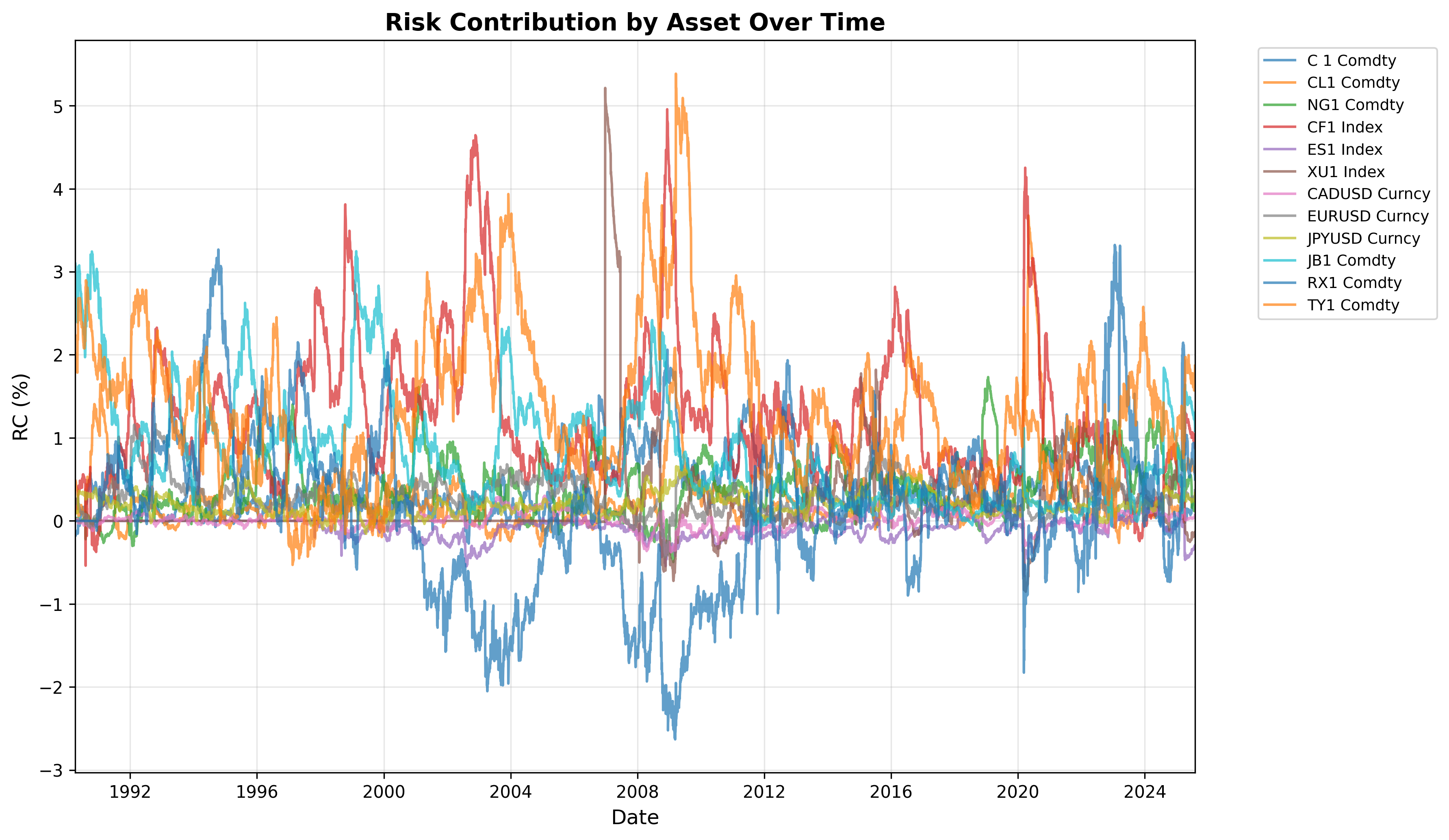}
		\caption{RC over time for each position in the long-short portfolio (Method 1: current weights, rolling covariance)}
		\label{fig:ls_rc_history}
	\end{figure}

	Figure~\ref{fig:ls_rc_history} shows LS portfolio RC over time for each position (Method 1: current weights, rolling covariance).
	
\begin{figure}[H]
		\centering
		\includegraphics[width=1\textwidth]{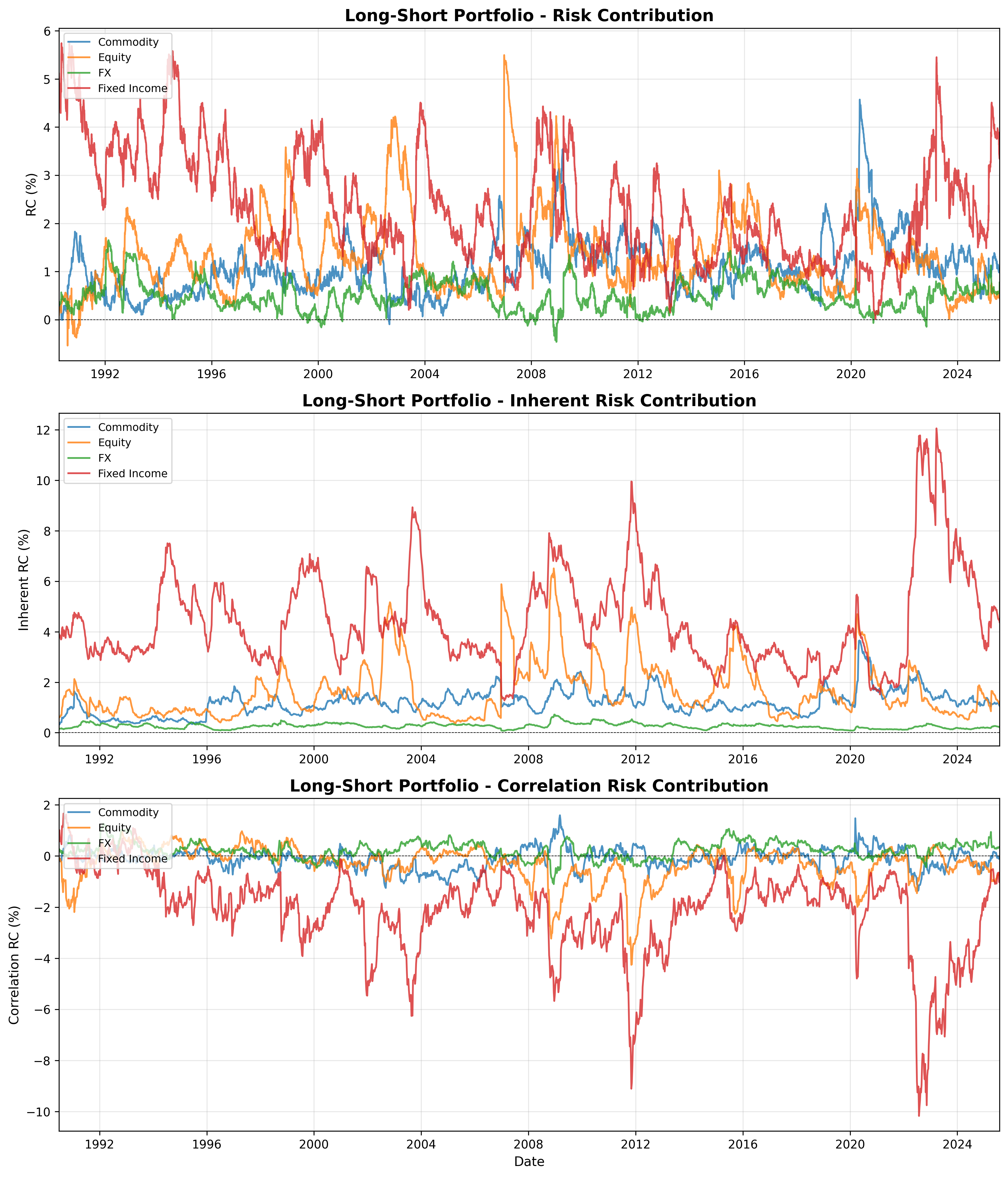}
		\caption{RC over time for the long-short portfolio aggregated by asset group, with total RC equal to portfolio volatility at each date.}
		\label{fig:ls_rc_history_asset_group}
	\end{figure}

	Figure~\ref{fig:ls_rc_history_asset_group} shows LS portfolio RC by asset group. Fixed income typically carries the largest inherent component, with the ordering of other groups varying over time.

	Table~\ref{tab:crisis_attr} provides a quantitative attribution of RC changes into inherent and correlation components for the LS portfolio across five selected crisis episodes. For each episode and the most-affected asset class, it reports the average inherent and correlation RC in the pre-crisis and crisis windows and the change between them. Equity inherent RC rises substantially ahead of the 2008 Global Finacial Crisis (GFC) ($\Delta\text{Inh} = {+}2.34$) while the correlation component changes only modestly ($\Delta\text{Corr} = {-}0.33$), indicating that the pre-GFC volatility build-up was driven primarily by stand-alone equity risk. During the GFC peak, equity inherent RC rises again ($\Delta\text{Inh} = {+}2.24$) while the correlation component falls sharply ($\Delta\text{Corr} = {-}1.61$): the LS portfolio carries a net-short equity position through the large short CF1 exposure, so equities provide greater diversification when equity markets decline. Commodities tell a different story, with both components rising ($\Delta\text{Inh} = {+}0.86$, $\Delta\text{Corr} = {+}0.41$), consistent with coordinated cross-asset liquidation, a pattern documented by Longin and Solnik (\cite{longinExtremeCorrelationInternational2001}). During the 2011--12 European Sovereign Debt Crisis, fixed-income inherent RC rises ($\Delta\text{Inh} = {+}2.47$) while the correlation component falls by nearly the same amount ($\Delta\text{Corr} = {-}2.34$), so total RC changes very little; rising bond volatility is largely offset by the flight-to-safety dynamic that makes long bond positions more negatively correlated with the rest of the portfolio. The Covid crash repeats the equity pattern seen in the GFC ($\Delta\text{Inh} = {+}2.12$, $\Delta\text{Corr} = {-}0.77$), while commodities again see both components rise ($\Delta\text{Inh} = {+}0.47$, $\Delta\text{Corr} = {+}0.44$). The 2022 Federal Reserve tightening cycle generates the largest fixed-income change in the sample: inherent RC rises by $+6.18$ as interest rate volatility surges globally, while the correlation component falls by $-5.12$, so the net change in total fixed-income RC is only $+1.07$; the LS portfolio's net-short equity position (via CF1) gained as equities declined during this period, remaining negatively correlated with falling bond prices and limiting the overall impact on portfolio risk. Fixed income's negative correlation component partially offsets its high inherent component throughout the sample. Although the portfolio is long-short, it is not risk-neutral, so it remains subject to directional risk during market crashes.
	
	\begin{table}[H]
		\centering
		\caption{Crisis Period RC Attribution for the LS Portfolio using Method 1 period averages with current LS weights applied to historical covariances (daily volatility percentages; $\Delta$ = crisis minus pre-crisis).}
		\label{tab:crisis_attr}
		\footnotesize
		\setlength{\tabcolsep}{3pt}
		\pgfplotstableset{
			col sep=comma,
			string type,
			columns/Crisis Period/.style={
				string type,
				column type={|l},
				column name={Crisis Period}
			},
			columns/Asset Class/.style={
				string type,
				column type={l|},
				column name={Asset Class}
			},
			columns/Pre Inh/.style={
				column type={S[table-format=-1.2, round-mode=places, round-precision=2]},
				column name={\multicolumn{1}{c}{Inh}}
			},
			columns/Pre Corr/.style={
				column type={S[table-format=-1.2, round-mode=places, round-precision=2]|},
				column name={\multicolumn{1}{c|}{Corr}}
			},
			columns/Crisis Inh/.style={
				column type={S[table-format=-1.2, round-mode=places, round-precision=2]},
				column name={\multicolumn{1}{c}{Inh}}
			},
			columns/Crisis Corr/.style={
				column type={S[table-format=-1.2, round-mode=places, round-precision=2]|},
				column name={\multicolumn{1}{c|}{Corr}}
			},
			columns/Delta Inh/.style={
				column type={S[table-format=-1.2, round-mode=places, round-precision=2]},
				column name={\multicolumn{1}{c}{$\Delta$Inh}}
			},
			columns/Delta Corr/.style={
				column type={S[table-format=-1.2, round-mode=places, round-precision=2]|},
				column name={\multicolumn{1}{c|}{$\Delta$Corr}}
			},
			every head row/.style={
				before row={\hline & & \multicolumn{2}{c|}{Pre-Crisis} & \multicolumn{2}{c|}{Crisis} & \multicolumn{2}{c|}{Change} \\ \hline},
				after row=\hline
			},
			every last row/.style={after row=\hline}
		}
		\begin{filecontents*}{tables/ls_crisis_attribution.csv}
Crisis Period,Asset Class,Pre Inh,Pre Corr,Crisis Inh,Crisis Corr,Delta Inh,Delta Corr
Pre-GFC (2007),Equity,0.9513712695847752,-0.08772049043605937,3.292906779973559,-0.41493866288049663,2.3415355103887836,-0.3272181724444373
GFC Peak (2008-09),Equity,2.6793192619408477,-0.5093132542929535,4.91833511323213,-2.118840279551253,2.2390158512912826,-1.6095270252582994
GFC Peak (2008-09),Commodity,0.8967868994173689,0.30206809487731834,1.757818259461966,0.7123116058109823,0.8610313600445971,0.41024351093366396
European Debt (2011-12),Fixed Income,4.34233422444342,-2.9074539895658926,6.815383232885713,-5.250684644410976,2.4730490084422927,-2.343230654845083
Covid (2020),Equity,0.9680422145832634,-0.4479719060100652,3.0876815811167,-1.2149700179827976,2.1196393665334368,-0.7669981119727324
Covid (2020),Commodity,1.2428329791664232,0.021915283171029906,1.70972501421598,0.4644903796595621,0.4668920350495569,0.44257509648853216
Fed Tightening (2022),Fixed Income,2.1437403914326816,-1.0389272320729313,8.327918430677483,-6.154311032976008,6.184178039244802,-5.115383800903077
\end{filecontents*}
\pgfplotstabletypeset{tables/ls_crisis_attribution.csv}
	\end{table}

\begin{figure}[H]
		\centering
		\includegraphics[width=1\textwidth]{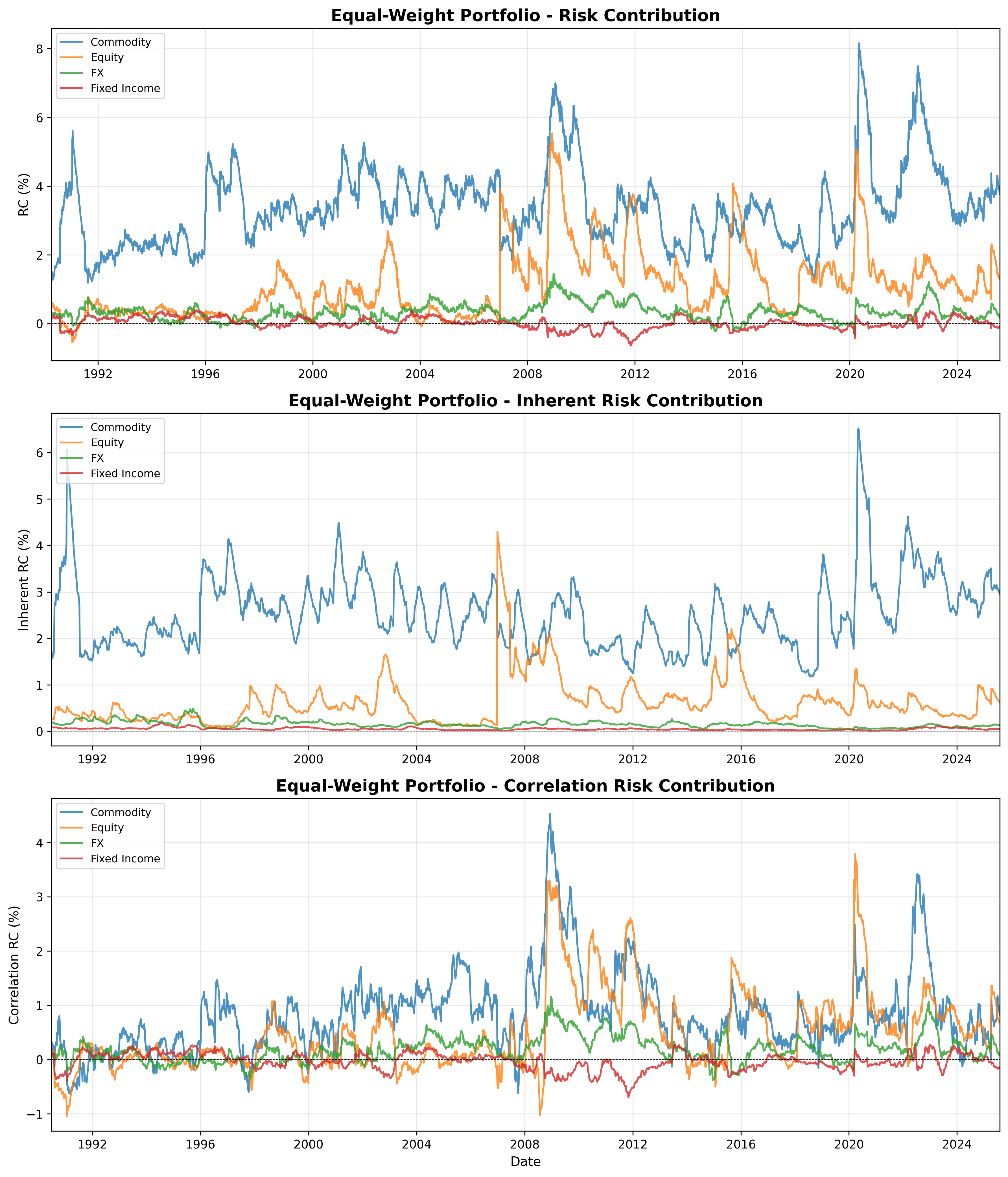}
		\caption{RC over time for the equal-weighted portfolio aggregated by asset group, with total RC equal to portfolio volatility at each date.}
		\label{fig:eq_rc_history_asset_group}
	\end{figure}

	Figure~\ref{fig:eq_rc_history_asset_group} shows EQ portfolio RC by asset group. Commodities and equities dominate. Equities show an inherent-led rise in 2007; GFC 2008 shows correlation-driven spikes in equities and commodities consistent with coordinated liquidation. Because the portfolio is fully long, equity correlation RC rises during sell-offs, unlike in the LS portfolio where the net-short position causes it to fall.
	
\begin{figure}[H]
		\centering
		\includegraphics[width=1\textwidth]{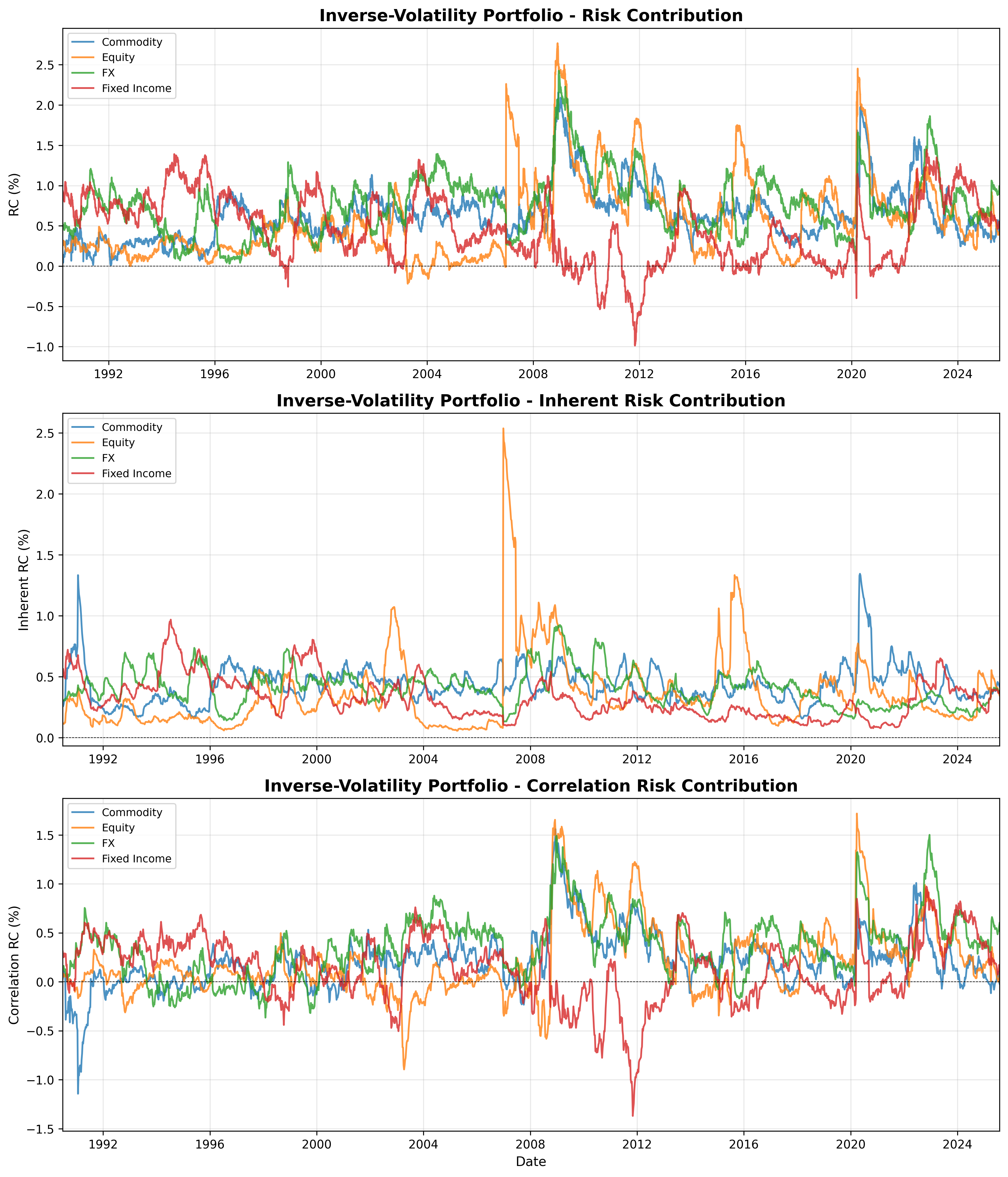}
		\caption{RC over time for the inverse-volatility portfolio aggregated by asset group, with total RC equal to portfolio volatility at each date.}
		\label{fig:rp_rc_history_asset_group}
	\end{figure}

	Figure~\ref{fig:rp_rc_history_asset_group} shows IV portfolio RC by asset group. Inverse-volatility weighting produces more even contributions than equal-weight, though not equal RC. Pre-GFC equity elevation is more prominent and FX plays a larger role during GFC and Covid periods. The 2011--12 European Sovereign Debt Crisis shows the same fixed-income offset (rising inherent, falling correlation) as in the LS portfolio.
	
	\section{Conclusion} \label{sec:conclusion}
	
	This paper introduces the Inherent and Correlation Decomposition (ICD) of Risk Contribution, deriving a leave-one-out representation of RC that splits it into an inherent term (position's own scaled variance) and a correlation term (covariance with the other holdings). Because the decomposition sits inside RC, it inherits strict additivity at any fixed granularity. The structure differs from the multiplicative X-sigma-rho decomposition of Menchero and Davis, which uses the position's correlation with the full portfolio; ICD uses covariance with the other holdings, giving qualitatively different diagnostic signals when a position is large relative to the portfolio. The contributions are the specific split, its interpretive framing, and an exact iVol identity yielding a closed-form discrete hedging condition that differs from the marginal $\text{RC}(a) < 0$ and can disagree with it for over-hedged or levered positions. Crisis-period attribution across five market regimes in Table~\ref{tab:crisis_attr} shows that the components shift differently: a risk rise driven primarily by stand-alone volatility calls for a different response than one driven by correlation breakdown. Granularity must be fixed, since the split depends on how positions are booked.
	
	The decomposition uses existing covariance estimates and supports time-varying analysis via the two methods presented. Robust covariance estimators may improve stability in high-dimensional or noisy settings. Future work could extend to dynamic allocation and machine-learning-based covariance estimation. By separating stand-alone variance from interaction variance within each RC value, ICD provides structural diagnostic information that aggregate RC does not reveal.
	
	\section*{Declarations of Interest}

	The authors report no conflicts of interest. The authors alone are responsible for the content and writing of the paper.

	\section*{Acknowledgements}

	The authors used Claude (Anthropic) for style editing only. All intellectual content, analysis, and conclusions are the authors' own.

	\pagebreak

	\printbibliography
	
	\pagebreak
	
	\begin{appendices}
		
		\section{iVol Non-additivity} \label{app:ivol_nonadditivity}
		Non-additivity of iVol is well known in the risk-budgeting literature. This appendix provides a self-contained proof that the bound $\sum_{a\in N} \text{iVol}(a) \le \sigma_p$ holds for all weight vectors under the leave-one-out representation used in this paper, where $\sigma_{p\backslash\{a\}} = \sqrt{\operatorname{Var}(r_p - w_a r_a)}$.

		\begin{align*}
			\sigma_p &\ge \sum_{a\in N} \text{iVol}(a).
		\end{align*}

		\noindent From the definition of iVol in eq.~\eqref{eq:ivol_def},

		\begin{align*}
			\sum_{a\in N} \text{iVol}(a) &= \sum_{a\in N}\left(\sigma_p - \sigma_{p\backslash\{a\}}\right) \\
			&= |N|\, \sigma_p - \sum_{a\in N} \sigma_{p\backslash\{a\}}.
		\end{align*}

		\noindent It therefore suffices to show $\sum_{a\in N} \sigma_{p\backslash\{a\}} \ge (|N|-1)\, \sigma_p$. Summing the return identities over $a$ and using $\sum_{a\in N} w_a r_a = r_p$,

		\begin{align*}
			\sum_{a\in N}\left(r_p - w_a r_a\right) &= |N|\, r_p - r_p = (|N|-1)\, r_p.
		\end{align*}

		\noindent The triangle inequality for standard deviations, which follows from the covariance matrix being positive semi-definite, gives

		\begin{align*}
			(|N|-1)\, \sigma_p = \sqrt{\operatorname{Var}\!\left(\sum_{a\in N}\left(r_p - w_a r_a\right)\right)} &\le \sum_{a\in N} \sqrt{\operatorname{Var}(r_p - w_a r_a)} = \sum_{a\in N} \sigma_{p\backslash\{a\}}.
		\end{align*}

		\noindent Substituting back,

		\begin{align*}
			\sum_{a\in N} \text{iVol}(a) &= |N|\, \sigma_p - \sum_{a\in N} \sigma_{p\backslash\{a\}} \\
			&\le |N|\, \sigma_p - (|N|-1)\, \sigma_p \\
			&= \sigma_p.
		\end{align*}

		\noindent The triangle inequality is strict for any non-degenerate portfolio, so the bound is strict outside degenerate cases. Because the proof uses only the triangle inequality and the return identity $\sum_{a\in N}(r_p - w_a r_a) = (|N|-1)\,r_p$, it requires no restriction on the signs or magnitudes of the weights and holds for portfolios with short or levered positions.

		\section{Convergence of RC and Its Components} \label{app:convergence}
\begin{figure}[H]
			\centering
			\includegraphics[width=1\textwidth]{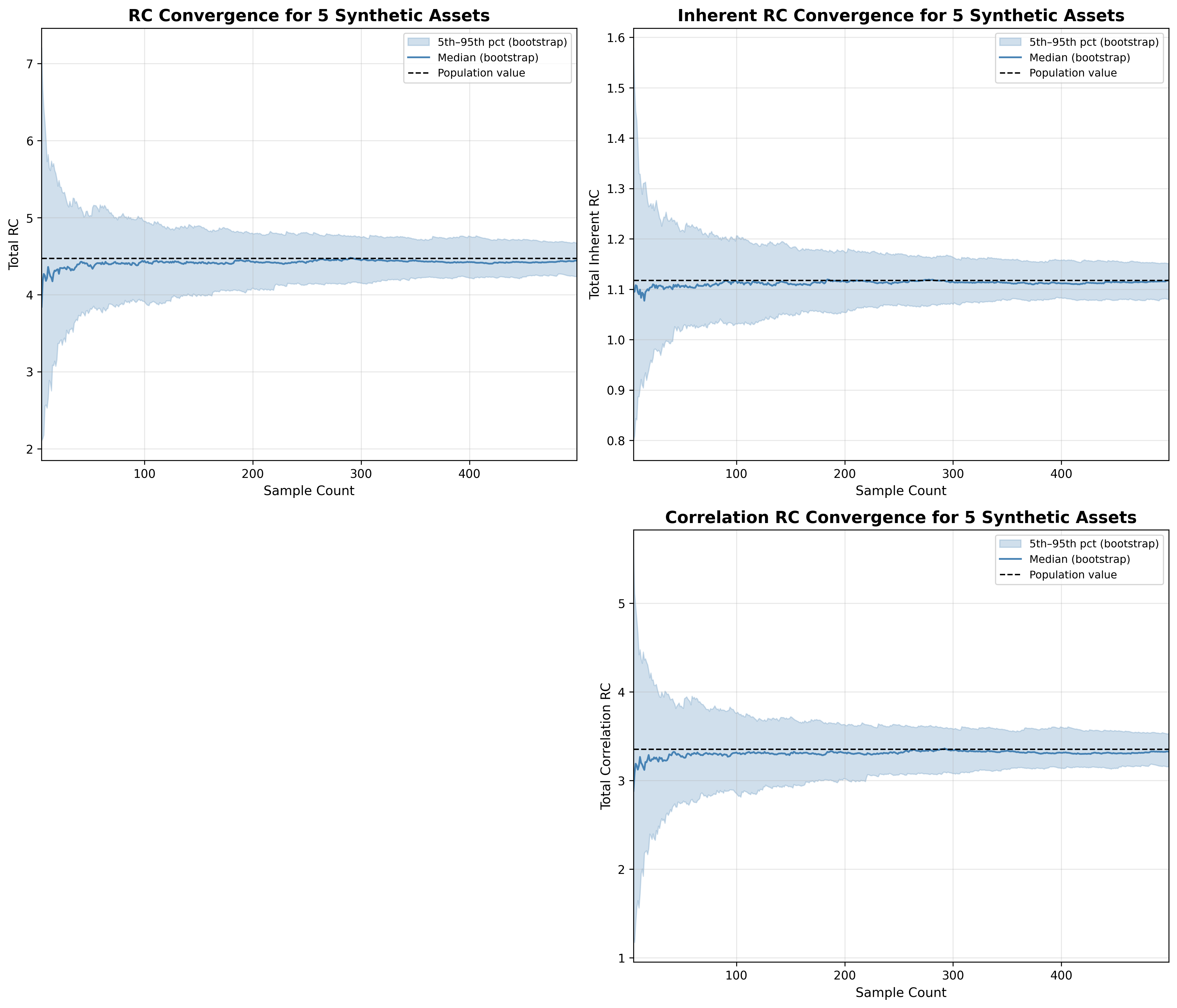}
			\caption{Convergence of RC and its components with five synthetic iid Gaussian assets (median and 5th--95th percentile bands across 100 replications; dashed lines show population targets)}
			\label{fig:rc_convergence}
		\end{figure}

		To examine point-estimator behavior of RC and its components, we simulate five synthetic assets with return vectors drawn from $\bm{r}_t \sim \mathcal{N}(\bm{0}, \bm{\Sigma})$ for $t \in [1, T]$. The covariance matrix $\bm{\Sigma}$ has diagonal values of 1 and off-diagonal values of 0.5. RC and its components are calculated with an expanding window starting at five days, so that the window length runs from $5$ up to $T$. The equal-weighted portfolio is used as the illustration. In this stylized setting, it coincides with both inverse-volatility and equal-risk-contribution weightings because the synthetic assets have identical volatility. Figure~\ref{fig:rc_convergence} shows the median and 5th-to-95th-percentile confidence bands for total RC, inherent RC, and correlation RC across 100 bootstrap replications at each window size, together with the population target as a reference line. All bands narrow substantially after approximately 125 observations. The convergence rates of the inherent and correlation components are similar to that of aggregate RC, suggesting the decomposition does not introduce additional estimation instability. This is a stylized i.i.d. Gaussian illustration; real portfolios face non-Gaussian returns and time-varying covariance.

	\end{appendices}
	
\end{document}